\documentclass[aps,draft]{revtex4}
\usepackage[dvips]{graphicx}
\begin{document}

\title{Weibel Instabilities in Dense Quantum Plasmas}
\author{ Levan N.Tsintsadze }
\affiliation{Department of Plasma Physics, Institute of Physics,
Tbilisi, Georgia}
\author{P. K.Shukla}
\affiliation{Institute f\"{u}r Theoretische Physik IV,
Ruhr-Universit\"{a}t Bochum, Bochum, Germany}

\date{\today}

\begin{abstract}
The quantum effect on the Weibel instability in an unmagnetized
plasma is presented. Our analysis shows that the quantum effect
tends to stabilize the Weibel instability in the hydrodynamic
regime, whereas it produces a new oscillatory instability in the
kinetic regime. A novel effect the quantum damping, which is
associated with the Landau damping, is disclosed. The new quantum
Weibel instability may be responsible for the generation of
non-stationary magnetic fields in compact astrophysical objects as
well as in the forthcoming intense laser-solid density plasma
experiments.
\end{abstract}


\maketitle

The Weibel instability \cite{wei} arises in variety of plasmas
including fusion plasmas, both magnetic and inertial confinement,
space/astrophysical plasmas, as well as in plasmas created by high
intensity free electron x-ray laser pulses. The Weibel instability
is of significant interest since it generates quasi-stationary
magnetic fields, which can account for seed magnetic fields in
laboratory \cite{est} and astrophysical plasmas \cite{med}. The
purely growing Weibel instability in a Maxwellian plasma is
excited by the anisotropy of the electron distribution function.
The linear and nonlinear aspects of the Weibel instability in
classical Maxwellian electron-ion plasmas are fully understood
\cite{davidson}.

However, in dense plasmas, such as those in compact astrophysical
objects (e.g. the interior of the white dwarfs, neutron stars/
magnetars, supernovae) as well as in the next generation intense
laser-solid density plasma experiments \cite{marklund}, in
nanowires and in micromechanical systems, one notices the
importance of quantum effects \cite{gardner} involving the
electron tunneling at nanoscales. In dense quantum plasmas, the de
Broglie wavelength associated with the plasma particles is
comparable to the interparticle spacing, and one uses either the
Wigner-Maxwell equations \cite{wigner} or quantum hydrodynamical
models \cite{manfredi} to investigate numerous collective
interactions \cite{marklund}. To study quantum effects in plasmas,
Klimontovich and Silin \cite{kli} derived a general kinetic
equation for the quantum plasma, and linearizing that equation
they obtained linear dispersion relations for transverse
electromagnetic (EM) waves, as well as for longitudinal waves. The
latter have also been studied by Pines \cite{pines}, who reported
the dispersion of electron plasma oscillations at quantum scales
due to the Bohm potential \cite{gardner} that causes electron
tunneling.

In this Letter, we present new aspects of the Weibel instability
in an unmagnetized quantum plasma. For our purposes, we  use the
dispersion relation $k^2c^2/\omega^2 = \varepsilon^{tr}$ for the
EM waves, with the following transverse dielectric permeability
\cite{kli,kuz}
\begin{eqnarray}
\label{per}
\varepsilon^{tr}=1-\sum_\alpha\frac{\omega_{p\alpha}^{2}}
{\omega^{2}}+ \sum_\alpha\frac{2\pi q_\alpha^{2}}{
\hbar\omega^{2}}\cdot\int\frac{d^{3}p}{\omega -\vec{k} \cdot
\vec{v}} v_\perp^{2}\left[ f_{0\alpha }\Bigl( \vec{p}+\frac{\hbar
\vec{k}}{2}\Bigr)-f_{0\alpha }\Bigl( \vec{p}-\frac{\hbar
\vec{k}}{2}\Bigr) \right]\ ,
\end{eqnarray}
where ${\bf k}$ is the wave-vector, $c$ is the speed of light in
vacuum, $\omega$ is the wave frequency, $\omega_{p\alpha}$ is the
plasma frequency of the particle species $\alpha$, $q_\alpha$ is
the charge, $\hbar$ is the Planck constant divided by $2\pi$,
$f_{0\alpha}$ is the equilibrium distribution function,
$\vec{p}_\alpha$ is the momentum.  In the non-relativistic limit,
we have $\vec{p}_\alpha=m_{0\alpha} \vec{v}$, where $m_{0\alpha}$
is the rest mass and $\vec{v}$ is the velocity vector. Using the
notations $\vec{v}+ \hbar\vec{k}/ 2m_{0\alpha} \rightarrow\vec{v}$
in the first integral, and $\vec{v}- \hbar\vec{k}/2m_{0\alpha}
\rightarrow\vec{v}$ in the second one, Eq. (\ref{per}) is
rewritten in the form
\begin{eqnarray}
\label{rnper}
\varepsilon^{tr}=1-\sum_\alpha\frac{\omega_{p_{\alpha
 }}^{2}}{\omega^{2}}
+\sum_\alpha\frac{2\pi q_{\alpha }^{2}}{ \hbar\omega^{2}}\cdot\int
d^3v
 v_\perp^2 f_{0\alpha}(v)\left(
\frac{1}{\omega +\frac{\hbar k^2}{2m_{0\alpha}} - \vec{k} \cdot
\vec{v}}-\frac{1}{\omega -\frac{\hbar k^2}{2m_{0\alpha}
 }-\vec{k} \cdot \vec{v}}\right)   \ .
\end{eqnarray}

Let us choose an anisotropic distribution function
\begin{eqnarray}
\label{dis} f_{0\alpha }=n_{0\alpha} A_\alpha
 \exp\left(-\frac{m_{0\alpha} v_\perp^2}
{2 T_{\alpha \perp}} -\frac{m_{0 \alpha} v_\parallel^2}{2T_{\alpha
 \parallel}}\right) \ ,
\end{eqnarray}
where $n_{0\alpha}$ is the equilibrium density, $T_{\alpha \perp}$
 $(T_{\alpha \parallel})$ is the
temperature transverse (parallel) to ${\bf k}$. The above
distribution function can also be expressed as
\begin{eqnarray}
\label{adis} f_{0\alpha }=n_{0\alpha} f_\alpha(v_\perp^2) \delta
(v_z) \ , \ or \ f_{0\alpha} =n_{0\alpha} f_\alpha(v_\perp^2)
\delta (v_z-u_{0z}) \ ,
\end{eqnarray}
where $u_{0z}$ is the equilibrium drift along the $z$ axis in a
Cartesian coordinate system.

Focusing on transverse EM waves propagating along the $z$ axis, we
can take $\vec{k} = (0,0,k)$ and $\vec{k} \cdot \vec{v}= kv_z$,
and introduce
\begin{eqnarray}
\label{gen} \frac{1}{2}m_{0\alpha} \int d^3v
v_\perp^2f_\alpha(v_\perp,v_z)=m_{0\alpha}n_{0\alpha}<\frac{v_\perp^2}{2}>
\int dv_z f_{0\alpha}(v_z) \ ,
\end{eqnarray}
to rewrite Eq. (\ref{rnper}) as
\begin{eqnarray}
\label{fper} \varepsilon^{tr}=1-\sum_\alpha\frac{\omega_{p_{\alpha
 }}^{2}}{\omega^{2}}
+\sum_\alpha\frac{\omega_{p_{\alpha}}^{2}}{\omega^{2}\hbar}\frac{m_{0\alpha}<v_\perp^2>}{2}\int
dv_zf_{0\alpha}(v_z)\left(\frac{1}{\omega_+ -
kv_z}-\frac{1}{\omega_- -kv_z}\right)   \ ,
\end{eqnarray}
where $\omega_\pm=\omega\pm \hbar k^2/ 2m_{\alpha0}$. We note that
in the case of the  Maxwellian distribution we would have $(1/2)
m_{0\alpha}n_{0\alpha}<v_\perp^2>= n_{0\alpha}T_{ \alpha\perp}$.

We now consider some special cases for an electron plasma, with
fixed ion background. First, choosing $f_{0\alpha}=\delta(v_z)$,
we obtain
\begin{eqnarray}
\label{fir}\frac{k^2c^2}{\omega^2}=1-\frac{\omega_{pe}^2}{\omega^2}
\left(1+\frac{ k^2<v_\perp^2>}{2(\omega^2-\eta^2)}\right) \ ,
\end{eqnarray}
where
$\eta = \hbar k^2/2m_{0e}$. Supposing  that $\omega^2\ll\eta^2$,
we obtain from (\ref{fir})
\begin{eqnarray}
\label{sfir} \omega^2=k^2c^2+\omega_p^2\left(1-\frac{
k^2<v_\perp^2>}{2\eta^2}\right).
\end{eqnarray}
Equation (\ref{sfir}) predicts a purely growing quantum
instability if $ 2 <v_\perp^2>m_{0e}^2 >\hbar^2k^2
 \left(1+k^2c^2/\omega_{pe}^2\right)$.
It should be stressed that if in the expression (\ref{fper})
$f_0(v_z)$ is the Maxwellian distribution function of the form
$f_0(v_z)=\left(m_{0e}/2\pi T_{e\parallel}\right)^{1/2}\exp\left(-
m_{0e}v_z^2/2T_{e\parallel}\right)$, then by assuming
$\left|\omega-kv_z \right| \ll \hbar k^2/2m_{0e}$ in the integral
of Eq.(\ref{fper}), one would obtain the dispersion relation
(\ref{sfir}), which shows that this expression does not depend on
the parallel electron temperature.

Equation(\ref{fir}) indicates that the quantum effect can
stabilize the Weibel instability for short wavelengths. We observe
that Eq. (\ref{fir}) has four roots, two of which are the low
frequencies ($\omega\ll\omega_{pe}$). From Eq.(\ref{fir}) we
obtain
\begin{eqnarray}
\label{lfr} \omega^2=\eta^2-\frac{\omega_{pe}^2k^2
 <v_\perp^2>}{2(\omega_{pe}^2+k^2c^2-\eta^2)}\ .
\end{eqnarray}
In order to estimate the wavelengths for which the quantum effect
can stabilize the Weibel instability, we suppose that
$\omega_{pe}^2\sim k^2c^2$. This leads to a condition of
stabilization from Eq.(\ref{lfr})
\begin{eqnarray*}
\frac{\hbar^2k^2}{m_{0e}^2} >  \ <v_\perp^2> \ {\rm or} \
\frac{T_{e\perp}}{2m_{0e}} \ .
\end{eqnarray*}

Next, we study the kinetic quantum effect in plasmas. In the
following, we assume that the distribution function $f_0(v_z)$ is
of the form
\begin{eqnarray}
\label{max}
f_0(v_z)=\frac{1}{v_\parallel\sqrt{\pi}}\exp\left(-\frac{v_z^2}{v_\parallel^2}\right)
 \ ,
\end{eqnarray}
where $v_\parallel=\left( 2 T_{e\parallel}/m_{0e}\right)^{1/2}$.
Introducing the dimensionless quantities
\begin{eqnarray*}
u=\frac{v_z}{v_\parallel} \ , \hspace{1cm}
z_\pm=\frac{\omega\pm\eta}{kv_\parallel}\ ,
\end{eqnarray*}
we can express Eq. (\ref{fper}) as
\begin{eqnarray}
\label{kper}
 \varepsilon^{tr}=1-\frac{\omega_{pe}^2}{\omega^2}+\frac{\omega
_p^2}{\omega^2}\frac{<\frac{m_{0e}v_\perp^2}{2}>}{\hbar
kv_\parallel} \frac{1}{\sqrt{\pi}}\int_{-\infty}^{+\infty}du
e^{-u^2}\left(\frac{1}{z_+-u}-\frac{1}{z_--u}\right) \ .
\end{eqnarray}
Here the integral $(1/\sqrt{\pi}) \int (z-u)^{-1} du \exp(-u^2) =
-i \sqrt{\pi}w(z)$, where
\begin{eqnarray}
\label{wz} w(z)=\exp(-z^2)\left(1+ \frac{2
 i}{\sqrt{\pi}}\int_0^z\exp(t^2)dt\right) \ .
\end{eqnarray}
The function $w(z)$ is related with the function $I_+(z)$ through
\begin{eqnarray}
\label{bwz} \frac{I_+(z)}{z}=- i \sqrt{\pi}w(z) \ ,
\end{eqnarray}
and the asymptotes of $I_+ (z)$  are
\begin{eqnarray}
\label{bas} I_+(z)=1+ \frac{1}{2z^2}+\frac{3}{z^4}+...- i
\sqrt{\pi}z \exp(-z^2) \ for \ \left|z\right| \gg 1 \ {\rm and} \
\left|{\rm Im} z \right| \ll
 1\ , \nonumber
\\
I_+(z)=- i \sqrt{\pi}z(1-z^2) +2z^2 \ {\rm for} \ \left|z\right|
\ll 1
 \ .
\end{eqnarray}

We now rewrite the expression (\ref{kper}) as
\begin{eqnarray}
\label{kperb} \varepsilon^{tr}=1 - \frac{\omega _{pe}^2}{\omega^2}
+\frac{\omega _{pe}^2}{2 \omega^2} \frac{m_{0e}<v_\perp^2>}{\hbar
 kv_\parallel}
\left(\frac{I_+(z_+)}{z_+} -\frac{I_+(z_-)}{z_-}\right) \ .
\end{eqnarray}

Consider the  case $z_\pm\gg 1$, so that Eq. (\ref{kperb}) can be
written as
\begin{eqnarray}
\label{dis1}\frac{k^2c^2}{\omega^2}=1-\frac{\omega_{pe}^2}{\omega^2}\left(1+\frac{
k^2 <v_\perp^2>}{2(\omega^2-\eta^2)}\right)+ 2 i
 \frac{\sqrt{\pi}\omega_{pe}^2}{\omega^2}
\frac{m_{0e} <v_\perp^2>} {2\hbar kv_\parallel}{\rm sinh} \left(
\frac{\hbar \omega}{2T_{e\parallel}}\right)\exp
 \left(-\frac{\omega^2+\eta^2}
{ k^2v_\parallel^2}\right) \ .
\end{eqnarray}
We specifically note here that if ${\rm
sinh}\Bigl(\frac{\hbar\omega}{2T_{e\parallel}}\Bigr)\gtrsim 1$,
then we get the result which we call the quantum damping. In the
opposite case, i.e., $\frac{\hbar\omega}{T_{e\parallel}}\ll 1$, we
obtain the classical damping.

In the low-frequency limit, viz. $\omega^2\ll\omega_{pe}^2$, Eq.
(\ref{dis1}) admits solutions of the form $\omega=\omega_r +
i\omega_i$, where the real and imaginary parts of the frequency
are given by, respectively,
\begin{eqnarray}
\label{rsol} \omega_r
=\left(\eta^2-\frac{\omega_{pe}^2}{2(\omega_{pe}^2+k^2c^2)}
 <v_\perp^2>  k^2\right)^{1/2},
\end{eqnarray}
and for the quantum Landau damping (QLD)
\begin{eqnarray}
\label{isol} \omega_i
 =-\sqrt{\pi}\left(\frac{\omega_{pe}^2}{\omega_{pe}^2+k^2c^2}\right)^2
\frac{m_{0e} <v_\perp^2>}{2\hbar kv_\parallel}
\frac{k^2<v_\perp^2>}{2\omega_r}{sinh}
 \left(\frac{\hbar\omega_r}{2T_{e\parallel}}\right)
\exp\left(-\frac{\omega^2+\eta^2}{k^2v_\parallel^2}\right) \ .
\end{eqnarray}

Next, for $z_\pm\ll 1$, we have $I_+(z)=- i
\sqrt{\pi}z(1-z^2)+2z^2$, and for the last term in Eq.
(\ref{kperb}) we obtain $z_{+}^{-1} I_+(z_+)- z_{-}^{-1} I_+(z_-)
=(2\eta/kv_\parallel)+2 i
 \sqrt{\pi}
(\eta/kv_\parallel)\omega/kv_\parallel$. In such an approximation,
we obtain
\begin{eqnarray*}
\varepsilon^{tr}=1-\frac{\omega_{pe}^2}{\omega^2}\left(1-
 \frac{m_{0e}<v_\perp^2>}{2 T_{e\parallel}}
\right)+ i \sqrt{\pi} \frac{m_{0e}<v_\perp^2>}{2
 T_{e\parallel}}\frac{\omega_{pe}^2}{\omega kv_\parallel} \ ,
\end{eqnarray*}
or
\begin{eqnarray}
\label{kin} k^2c^2+\omega_{pe}^2\left(1-\frac{m_{0e}<v_\perp^2>}{2
T_{e\parallel}}\right)-i \sqrt{\pi}
 \frac{m_{0e}<v_\perp^2>}{2T_{e\parallel}}\frac{\omega_{pe}^2\omega}{ kv_\parallel}=0 \ ,
\end{eqnarray}
which admits the solution
\begin{eqnarray}
\label{sol2} \omega=- i \frac{2}{\sqrt{\pi}}
\frac{T_{e\parallel}}{m_{0e}<v_\perp^2>} \frac{
kv_\parallel}{\omega_{pe}^2}\left[k^2c^2+\omega_{pe}^2
\left(1-\frac{m_{0e}<v_\perp^2>}{2T_{e\parallel}}\right)\right] \
.
\end{eqnarray}
Equation (\ref{sol2}) admits a purely growing instability if
\begin{eqnarray}
\label{con} \frac{m_{0e}<v_\perp^2>}{2
 T_{e\parallel}}>\frac{k^2c^2+\omega_{pe}^2}{\omega_{pe}^2}.
\end{eqnarray}

Finally, we consider the range of frequencies $\omega_+\gg
kv_z\gg\omega_- \ $ (Figure 1), or $\left|z_+\right| \gg 1$ and
$\left| z_-\right| \ll 1$. Clearly such situation can be realized
in a quantum case alone. In this case, Eq. (\ref{kperb}) reduces
to
\begin{eqnarray}
\label{fin} \varepsilon^{tr}=1-\frac{\omega
_{pe}^2}{\omega^2}+\frac{\omega_{pe}^2}{\omega^2}\frac{m_{0e}
 <v_\perp^2>}{2\hbar
kv_\parallel}\left(\frac{1}{z_+}-2z_-+ i \sqrt{\pi}\right) \ .
\end{eqnarray}
which yields, in the first approximation,
\begin{eqnarray}
\label{finap} \varepsilon^{tr}=1-\frac{\omega _{pe}^2}{\omega^2}
\left(1- i \sqrt{\pi}\frac{m_{0e} <v_\perp^2>}{2\hbar
 kv_\parallel}\right) \ .
\end{eqnarray}
Accordingly, in this case, the dispersion relation reads
\begin{eqnarray}
\label{fdis} \omega^2=\omega_{pe}^2\left(1- i
\sqrt{\pi}\frac{m_{0e}<v_\perp^2>} {2 \hbar
kv_\parallel}\right)+k^2c^2 \ .
\end{eqnarray}

As is well known, the classical Weibel instability is a purely
growing instability. We now show that the quantum effect leads to
a new type of Weibel instability, which we refer to as the Weibel
oscillatory instability. To this end, we rewrite Eq. (\ref{fdis})
as
\begin{eqnarray}
\label{nfdis}
\omega=\pm\sqrt{\omega_{pe}^2+k^2c^2}(1+Q^2)^{1/4}\left(
\cos\frac{\varphi}{2}- i \sin\frac{\varphi}{2}\right) \ ,
\end{eqnarray}
where $\varphi= {\rm arctg} Q$ and
$Q=\frac{\sqrt{\pi}\omega_{pe}^2}{(\omega_{pe}^2+k^2c^2)}
 \frac{m_{0e}<v_\perp^2>}
{2\hbar kv_\parallel}$.

Let us consider two cases. First, for $Q\ll 1$ and $\varphi\sim Q$
the real and imaginary parts of the frequencies, deduced from
Eq.(\ref{nfdis}), are
\begin{eqnarray}
\label{fsolr} \omega_r\approx \pm\sqrt{\omega_{pe}^2+k^2c^2} \ ,
\end{eqnarray}
\begin{eqnarray}
\label{fsoli} \omega_i=\pm
 \frac{\sqrt{\pi}\omega_{pe}^2}{\sqrt{\omega_{pe}^2+k^2c^2}}
\frac{m_{0e}<v_\perp^2>}{2\hbar kv_\parallel} \ .
\end{eqnarray}

More vigorous effect is obtained when $Q\gg 1$. Namely, in this
case $\varphi\sim\frac{\pi}{2}$, and we have
\begin{eqnarray}
\label{vig} \omega_r = \omega_i = 0.7\sqrt{\omega_{pe}^2+k^2c^2}
 Q^{1/2} \ .
\end{eqnarray}

In summary, we have investigated the quantum mechanical effects on
the Weibel instability in an unmagnetized plasma containing
electron energy anisotropy. It is shown that the quantum effect
stabilizes the Weibel instabilities, but a new type of Weibel
instabilities, the quantum Weibel instabilities are found. These
instabilities describe the quantum wave excitation with slow
damping by the quantum Landau mechanism. We have demonstrated
possibility of a novel oscillatory Weibel instability. The latter
may be responsible for the generation of non-stationary magnetic
fields in dense astrophysical objects, as well as in the next
generation intense laser-solid density plasma experiments. The
random walk of electrons in nonstationary magnetic fields can
produce anomalous electron transport at quantum scales in dense
plasmas.

\begin{figure}
\caption{Ranges of interaction of electrons with a wave for a
distribution over velocities }
\end{figure}

\begin{references}
\bibitem{wei} E.S.Weibel, Phys. Rev. Lett. {\bf 2}, 83 (1959).
\bibitem{est} K.Estabrook, Phys. Rev. Lett. {\bf 41}, 1808 (1978).
\bibitem{med} M.V.Medvedev and A.Loeb, Astrophys. J. {\bf 526},
697 (1999); R.Schlickeiser and P.K.Shukla, ibid, {\bf 599}, L57
(2003); K.-I.Nishikawa, R.Hardee, G.Richardson, R.Preece, H. Sol,
and G.J.Fishman, ibid, {\bf 622}, 927 (2005).
\bibitem{davidson} R.C.Davidson, D.Hammer, I.Haber, and C.E.Wagner,
Phys. Fluids {\bf 15}, 317 (1972).
\bibitem{marklund} M.Marklund and P.K.Shukla, Rev. Mod. Phys. {\bf
 78}, 581 (2006).
\bibitem{gardner} C.L.Gardner and C.Ringhofer, Phys. Rev. E {\bf
 53}, 157 (1996).
\bibitem{wigner} E.P.Wigner, Phys. Rev. {\bf 40}, 749 (1932).
\bibitem{manfredi} G.Manfredi and F.Haas, Phys. Rev. B {\bf 64},
 075316 (2001); G.Manfredi, Fields Inst. Comm. {\bf 46}, 263 (2005); P. K. Shukla
and B.Eliasson, Phys. Rev. Lett. {\bf 96}, 245001 (2006).
\bibitem{kli} Yu.L.Klimontovich and V.P.Silin, Zh. Eksp. Teor. Fiz.
{\bf 23} 151 (1952).
\bibitem{pines} D.Pines, J. Nucl. Energy C: Plasma Phys. {\bf 2}, 5 (1961).
\bibitem{kuz} M.V.Kuzelev and A.A.Rukhadze, Phys. Usp. {\bf 42} 603 (1999).
\end{references}
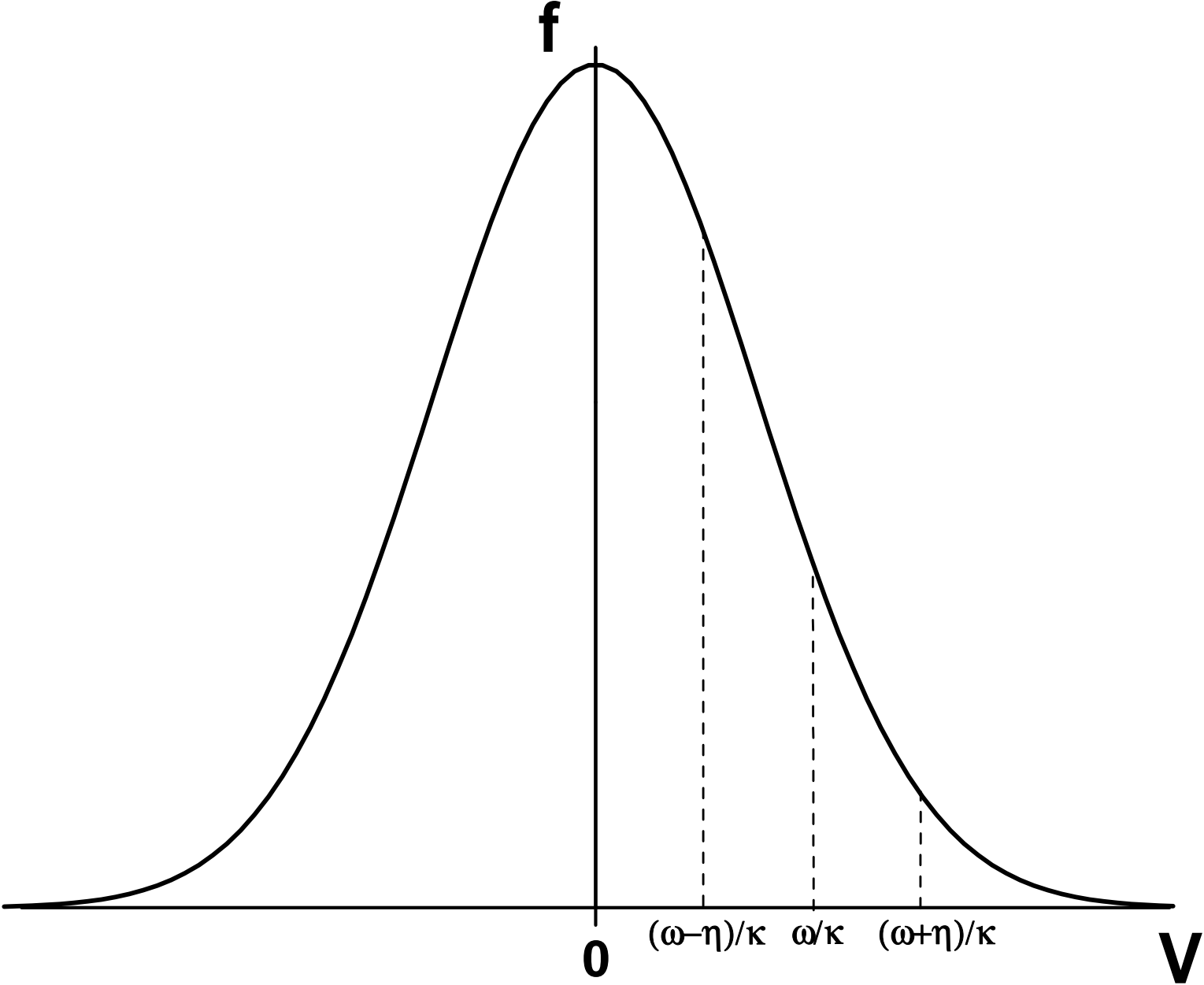
\end{document}